\documentstyle[eqsecnum,aps,twocolumn,prb]{revtex}
\newsavebox{\cdg}
\savebox{\cdg}(0,0)[bl]
{
\thicklines
\put(-0.5,-0.5){\line(1,0){1}}
\put(-0.5,-0.5){\line(0,1){1}}
\put(-0.5,0.5){\line(1,0){1}}
\put(0.5,-0.5){\line(0,1){1}}
}

\def\break#1{\pagebreak \vspace*{#1}}
\def\epsfig#1#2#3#4
         {
         \epsfysize=#2 \vbox{ \hglue#3 \epsfbox[#4]{#1} }
         }
\def\epsfigrot#1#2#3#4
         {
         \epsfxsize=#2 
         \setbox\rotbox=\hbox to #2{\epsfbox[#4]{#1}}
         \vbox{\hglue#3 \rotl\rotbox}
         }
\newbox\rotbox
\input rotate

\begin{document}
\draft
\title{Low-temperature nonequilibrium
transport in a Luttinger liquid}
\author{Ulrich Weiss$^1$, Reinhold Egger$^2$ and Maura Sassetti$^3$}
\address{
${}^1$Institut f\"ur Theoretische Physik, Universit\"at Stuttgart,
 D-70550 Stuttgart, Germany\\
${}^2$Fakult\"at f\"ur Physik, Albert-Ludwigs-Universit\"at,
D-79104 Freiburg, Germany\\
${}^3$Istituto di Fisica di Ingegneria,
Universit\`{a} di Genova, I-16146 Genova, Italy}
\maketitle
\widetext
\begin{abstract}
The temperature-dependent nonlinear conductance for transport
of a Luttinger liquid through a barrier is calculated in the
nonperturbative regime for $g=1/2 - \epsilon$, where $g$ is the
dimensionless interaction constant.  To describe
the low-energy behavior, we perform a leading-log summation
of all diagrams contributing to the conductance which is
valid for $|\epsilon| \ll 1$. With increasing external
voltage, the asymptotic low-temperature behavior displays a turnover from
the $T^{2/g-2}$ to a universal $T^2$ law.
\end{abstract}


\narrowtext

\section{INTRODUCTION}

The interplay between electron-electron interactions and disorder
has attracted a great deal of recent interest in condensed-matter
theory. Many-body correlations play an essential role
in quasi-one-dimensional (1D) fermionic systems, where
the usual Fermi liquid behavior is destroyed by the
interactions. The generic behavior of many interacting 1D
fermion systems is instead described in terms of the
Luttinger liquid model.\cite{lutt,haldane} One may then ponder
how transport in such a 1D ``quantum wire''
is affected by the presence of impurities
or tunnel barriers. A quantitative answer to this
 question is of immediate interest for several experimental
setups, e.g.~the tunneling of edge state excitations in the
fractional quantum Hall (FQH) regime,\cite{wen,webb} or
transport in narrow high-mobility heterostructure
channels.\cite{timp,tarucha}

The effect of one or a few barriers on the conductance of a Luttinger
liquid has first been studied by Kane and Fisher (KF).\cite{kane}
The Luttinger liquid model captures the low-energy properties of
correlated electron transport if two conditions are met:
one has short-ranged electron-electron interactions
(i.e.~the long-range $1/r$ tail of the Coulomb potential is
screened by nearby gates),
 and electron-electron
backscattering (BS) merely constitutes an irrelevant perturbation.
\cite{solyom} For the spinless case, BS is the exchange process of
 forward scattering, and therefore it is included readily by a
redefinition of the interaction constant.
All effects of the electron-electron
interaction are then encompassed in a single dimensionless
parameter $g$, which is approximately given
by $g\approx(1+U/2E_{\rm F})^{-1/2}$. Here, $U$ measures the
interaction strength and $E_{\rm F}$ is the Fermi energy.
We will consider the case of repulsive
 interactions, where one has $g<1$. The
noninteracting Fermi liquid case is recovered for $g=1$.
An ideal realization of this model is found in the
tunneling of edge state excitations in the FQH regime.
As shown by Wen,\cite{wen} these excitations are described
by a (chiral) Luttinger liquid with $g=\nu$, where $\nu$ is
\break{1.30in}
the filling factor. Since $\nu$ is a topological quantity
controlled by the bulk properties of the FQH bar, this provides
an excellent experimental testing ground.\cite{webb}
In addition, recent transport experiments in
1D quantum wires \cite{tarucha} have revealed Luttinger liquid
behaviors directly for correlated electrons.

It has been demonstrated that the presence of
an impurity potential term (the simplest realization of
disorder) leads to a metal-isolator quantum
phase transition at $T=0$ when going
from attractive to repulsive electron-electron interactions.\cite{kane}
For $g<1$, any (whatsoever weak) barrier will therefore result in zero
linear conductance, $G=0$.
One is then concerned with the low-energy properties of
the conductance, i.e.~the behavior of $G(T,V)$ in the limit
$T, eV/k_{\rm B}\ll T_{\rm K}$, where the
Kondo temperature
$T_{\rm K}$ sets the appropriate scale (see below).
As indicated by the RG flow, this is a nonperturbative
problem for weak barriers, and the past few years
have shown considerable effort being devoted to this
problem.\cite{wen,kane,glaz,glaz2,fend,moon}

By using a simple scaling argument, Kane and Fisher argued that the
linear ($V\to 0$) conductance should behave as $T^{2/g-2}$ in the asymptotic
low-temperature regime.\cite{kane} They obtained
this law from matching the RG flow in the nonperturbative
regime onto the large-barrier perturbative flow, thereby
exploiting universality arguments.
On the other hand, any finite  voltage leads
to a $g$-independent (universal) $T^2$ power law
instead of the $T^{2/g-2}$ behavior. As one can never be sure about
the validity of scaling arguments, we believe it is of use to
have exact results for the {\em nonequilibrium
finite-temperature} transport
as provided here.  We note that the exact solution
for the  case
$g=1/2$ has already been given within the bosonized
description.\cite{weiss} The linear conductance for $g=1/2$ has
also been calculated within
an equivalent fermionic model by Kane and Fisher,\cite{kane} who extended an
earlier zero-temperature approach by Guinea \cite{guinea} to finite
temperatures. Since the laws $T^{2/g - 2}$ and $T^2$ coincide for $g=1/2$,
however, it is crucial to consider $g\neq 1/2$.
In this paper, we start out from a dynamical approach and
present a nonperturbative leading-log summation for
$G(T,V)$ in the range $g=1/2-\epsilon$ with $|\epsilon|\ll 1$.
We systematically calculate the conductance in form of the series
\begin{equation}
G(T,V) = \sum_j a_j \epsilon^j_{} \ln^j_{} [{\rm max}(k_{\rm B}T,eV)]\;.
\end{equation}

Our study is related to recent work by Matveev, Yue and
Glazman.\cite{glaz} They have carried out a
leading-log summation for weak electron-electron
interactions. For the Luttinger liquid, this
would correspond to the case $g=1-\epsilon$ with $|\epsilon|\ll 1$.
Since the Kondo temperature vanishes with
an essential singularity as $g$ approaches 1 from below,
it is  desirable to have similar nonperturbative
results for stronger electron-electron interactions.
Their technique relies crucially on the assumption of weak interactions
and can therefore not be employed directly in other
regions of parameter space.

Basically parallel to our work, a powerful approach based
on the thermodynamic Bethe ansatz has been put forward
by Fendley, Ludwig and Saleur \cite{fend}
who studied the linear conductance at $g=1/3$.
They mapped the original boson model of the Luttinger liquid onto a
massless odd boson model which is integrable.\cite{ghoshal}  A basis of
massless fermion-type quasiparticles can be derived, in which the $S$
matrix for impurity scattering can be calculated exactly. The distribution
function and the density of states of these quasiparticles are given in terms
of integral equations with kernels determined by the completely elastic and
factorizable scattering matrix of the bulk. Expressed in terms of these
quantities, a Boltzmann-type rate expression for the conductance may then be
derived. Switching to a suitable momentum integral representation, our
solution for the current obtained from a dynamical approach
can be transformed such that we can read off the
$S$ matrix for impurity scattering, the pseudoenergy and the density of states
of the quasiparticles for $g=1/2 - \epsilon$ and finite voltage.
Therefore, our dynamical approach provides an independent check for results 
obtained by the thermodynamic Bethe ansatz, and furthermore gives
a nontrivial prediction for the density of states.

The outline of this paper is as follows. In Sec.~II, we present
the model for our study and give the formal solution.
The general diagrammatic expansion is explained in Sec.~III,
and the leading-log summation based on this diagrammatic expansion
is discussed in great detail in Sec.~IV. Results for the asymptotic
low-energy behavior as well as a conjecture for the
exact solution are presented in Sec.~V, followed by
concluding remarks in Sec.~VI. Some technical details of the
leading-log calculations are given in the Appendix.

\section{GENERAL FORMALISM}

To describe the problem of correlated electron transport,
we employ the standard bosonization method.\cite{lutt,haldane,kane}
The fermion field operator is expressed in terms of
two bosonic fields $\phi(x)$ and $\theta(x)$,
\[
\psi^\dagger(x) \sim \sum_{n\; {\rm odd}}
\exp[in(\sqrt{\pi}\theta(x)+k_{\rm F} x) + i\sqrt{\pi}\phi(x)]\;,
\]
where the boson fields obey the commutation relation
\[
[\phi(x),\theta(x')]=-(i/2)\,{\rm sgn}(x-x')\;.
\]
Taking short-range electron-electron interactions and neglecting
backscattering, this leads to the generic Luttinger liquid Hamiltonian
(we put $\hbar=1$)
 \begin{equation}
H_0=  \frac{v}{2} \int dx \left[ g(\partial_x
 \phi)^2 +  g^{-1} (\partial_x \theta)^2 \right] \;,
 \end{equation}
where $g$ is the dimensionless interaction constant and
$v$ denotes the sound velocity. The Fermi velocity is
then given by $v_{\rm F}=gv$. For the physical case of
repulsive interactions, one has $g<1$.
Although this model is universal, actual computations,
e.g.~of correlation functions, necessitate introduction of
a non-universal cutoff parameter $\omega_{\rm c}$
which is related to the bandwidth of the
non-interacting problem. In the end, since $H_0$ is
quadratic, everything can be solved exactly.\cite{lutt}

To model the impurity, we follow KF \cite{kane} and
consider a short-ranged scattering potential centered
at $x=0$. Omitting irrelevant multi-electron backscattering
processes, the important contributions are captured by
the $2k_{\rm F}$-component $V_0$ of this scattering
potential, and one has
 \begin{equation}\label{lutt-mod}
 H= H_0 + V_0  \cos[2\sqrt{\pi} \theta(0)] + eV\,\theta(0)/\sqrt{\pi}\;.
 \end{equation}
We have also included an external voltage $V$ by assuming
the voltage drop to occur at the barrier.
This is well justified for not too small barriers.

The model (\ref{lutt-mod}) describes scattering by an impurity potential
with effective strength $V_0$. Interestingly, results for a very high barrier
$V_0\gg \omega_{\rm c}$ can be obtained via an exact
 duality\cite{schmid} from the case $V_0\ll\omega_{\rm c}$.
For $g<1$, the large-barrier
problem can be treated in lowest order in the number of tunneling
transitions for any $T$ and $V$, whereas in the weak-barrier case
perturbation theory in $V_0$ is not possible at sufficiently low
temperature and small voltage.
Our subsequent treatment deals with
the interesting nonperturbative regime.  Putting $g=1/2-\epsilon$,
our results presented in Sec.~V hold under the conditions
\begin{equation} \label{conditions}
V_0/\omega_{\rm c}\ll 1\;,\quad
|\epsilon| \ll 1 \;, \quad eV/k_{\rm B} T_{\rm K} \ll 1 \;,
\quad T/T_{\rm K} \ll 1 \;,
\end{equation}
where  $T_{\rm K}$ is the Kondo temperature
\begin{equation}\label{kondo}
  T_{\rm K} = (\sqrt{\pi}V_0/\omega_{\rm c})^{g/(1-g)}
\sqrt{\pi} V_0/k_{\rm B}^{} \, 
\end{equation}
taken at $g=1/2-\epsilon$.  The Kondo temperature
provides an estimate for the extent of the nonperturbative regime. Note
that
$T_{\rm K}\to 0$ for $g\to 1^-$ as alluded to in the Introduction.

In the following, we study the nonlinear static conductance
$G(V,T) = \partial I/\partial V$.
From the bosonized form of the current operator, it takes the form
\begin{equation}
G(V,T) =(e/\sqrt{\pi})\; (\partial/\partial V)  \lim_{t\rightarrow \infty}
        \langle \dot{\theta}(0,t) \rangle \; , \label{conduc}
\end{equation}
where $\langle\cdots\rangle$ denotes the thermal average over
all modes of (\ref{lutt-mod}) away from $x=0$.
Remarkably, the model (\ref{lutt-mod}) is identical to the problem of a
Brownian particle moving in a periodic potential.\cite{weiss,guinea,schmid}
This can be directly shown by a unitary transformation,\cite{guinea}
and we exploit such a mapping here. The $\theta$-field taken
at the location of the impurity, $\theta(0)$, corresponds
to the position operator of the Brownian particle, and the
Ohmic dissipation is provided by the plasmon modes away from the barrier.
Thereby the conductance can be expressed in terms of
the nonlinear mobility of the Brownian
particle which is coupled to an Ohmic heat bath.

An exact formal expression for the conductance can be obtained
from a real-time path-integral  approach due to Feynman and
Vernon.\cite{fv} Employing the unitary transformation of Eq.~(\ref{lutt-mod})
onto the tight-binding limit of Brownian motion in a periodic potential,
one can write this expression in a ``Coulomb gas'' representation
of interacting discrete charges.\cite{kane} Alternatively,
our expression (\ref{kinkgas}) with (\ref{udef}) can be found by expanding the
impurity propagator in terms of auxiliary  charge
paths.\cite{schmid} The $\theta(0)$ paths can
then be eliminated analytically and one has to sum over
Hubbard-Stratonovich paths instead. These auxiliary paths
can be  parametrized by  the discrete charges  $\xi_j=\pm 1$.

In the end, the exact formal expression for the conductance
takes the form of a power series in $V_0^2$,
\begin{equation} \label{kinkgas}
G(V,T)/G_0 = 1 - (2\pi/eV) \,\mbox{Im}\, U(V,T)\;,
\end{equation}
where $G_0=g e^2/h$ and
$U$ describes the interacting charge gas
\begin{eqnarray}
U(V,T) &= &\sum_{m=1}^\infty  (iV_0)^{2m}
 \int\limits_0^\infty\! d\tau^{}_1\,d\tau^{}_2\cdots d\tau^{}_{2m-1}
\label{udef}\\
 &\times& \sum_{\{\xi \}}
\,W_m \,\prod_{j=1}^{2m-1}
e^{-igeV p_{j,m} \tau_j} \,\sin(\pi  p_{j,m} g) \;, \nonumber
\end{eqnarray}
with the interaction factor
\begin{equation}\label{intfac}
W_m = \exp\left( \sum_{j>k=1}^{2m} \xi_j
S(\tau_{jk}) \xi_k \right) \; .
\end{equation}
The interaction potential between the charges is given by
\begin{equation}
S(\tau) = 2g \ln[ (\omega_{\rm c}/\pi k_{\rm B} T)
\sinh(\pi k_{\rm B} T \tau) ]\;.
\end{equation}
The $2m-1$ integration times $\tau_j$ in order $V_0^{2m}$ are the
intermediate times between the $2m$ successive charges, and
$\tau_{jk}$ gives the time interval between charges $\xi_j$ and
$\xi_k$.
Charge configurations $\{\xi\}$ contributing to (\ref{udef}) are restricted as
follows. (1) In every order $m$, the $2m$ charges have zero total
charge, $\sum_{j=1}^{2m} \xi_j = 0$ (neutrality condition).
(2) The cumulative charge quantities $p_{j,m}$
\[
p_{j,m} = \sum_{i=j+1}^{2m} \xi_i = - \sum_{i=1}^{j} \xi_i
\]
can never be zero since otherwise the
corresponding phase factor $\sin(\pi p_{j,m} g)$ would vanish. Expressed in
graphical terms, there are no diagrams that can be subdivided into
disjoint parts each of them representing a neutral configuration.
(3) By convention, the first charge is $\xi_1=-1$. With that,
all $p_{j,m}$ are positive integers.

\begin{figure}
\thicklines
\begin{picture}(14,12)
\put(0.5,11){(a)}
\put(0,9.5){\line(1,0){13}}
\put(2,7.5){\line(0,1){2.0}}
\put(5,7.5){\line(0,1){2.0}}
\put(8,9.5){\line(0,1){2.0}}
\put(11,9.5){\line(0,1){2.0}}

\put(0.5,6){(b)}
\put(0,5){\line(1,0){13}}
\put(0,3){\line(1,0){13}}
\put(0,1){\line(1,0){13}}
\put(0.5,1){\circle*{0.32}}
\put(3.5,3){\circle*{0.32}}
\put(6.5,5){\circle*{0.32}}
\put(9.5,3){\circle*{0.32}}
\put(12.5,1){\circle*{0.32}}
\put(13.2,1){$p=0$}
\put(13.2,3){$p=1$}
\put(13.2,5){$p=2$}
\end{picture}
\caption[]{\label{fig1} Charge configuration $(--++)$
contributing in order $m=2$ to the expression
(\ref{udef}). (a) Diagram representing this
configuration. Time flows along
the horizontal line, vertical lines symbolize charges
$\xi=\pm 1$. (b) Accumulated charge $p_i=\sum_{j>i}\xi_j$ during
the course of the path. Circles stand for the
momentary position of the path.  For the shown configuration, an
 overall phase factor $2\pi\epsilon$
is acquired according to Eq.~(\ref{phase}). }
\end{figure}
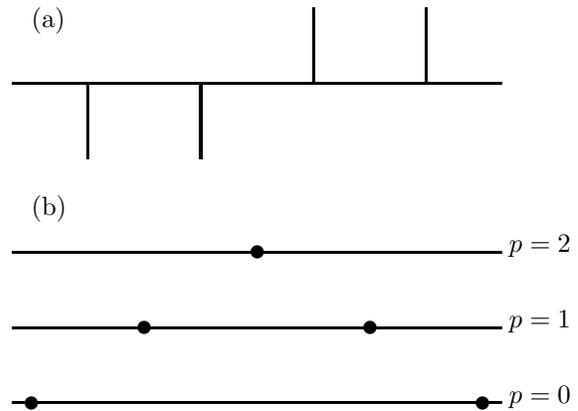

To illustrate these rules and our subsequent diagrammatic
approach, Fig.~\ref{fig1} shows a charge configuration
for $m=2$ which does contribute to Eq.~(\ref{udef}).
Putting $g=1/2-\epsilon$, the phase factors appearing in
Eq.~(\ref{udef}) can be written as
\begin{equation} \label{phase}
\sin(\pi p g) \simeq \left\{ \begin{array}{c@{\quad,\quad}l}
(-1)^{(p-1)/2}     & p\; {\rm odd,} \\
 \pi \epsilon p \,(-1)^{(p-2)/2} & p\; {\rm even .}
\end{array} \right.
\end{equation}
As one moves away from $p=0$, one has
to pay the phase factors $1, 2\pi \epsilon, -1, -4\pi\epsilon,\ldots$
for dwelling on states labelled $p=1,2,3,4,\ldots$,
respectively. This is displayed in Fig.~\ref{fig1}(b).
The charge configuration in Fig.~\ref{fig1} has a phase
prefactor $2\pi\epsilon$. On the other hand, the breathing
mode integral over the interior dipole [the $\tau_2$-integration of the
$(m\!=\!2)$--term in Eq.~(\ref{udef})] has a $1/\epsilon$ singularity due
to the $\tau^{-1+2\epsilon}$ short-time behavior of the intradipole
interaction $\exp[-S(\tau)]$. Hence the combined expression is well defined
even when $\epsilon=0$.

The essential scale which separates the nonperturbative from the perturbative
regime may be derived by using an adiabatic renormalization
scheme.\cite{leggett87}  Rather than working in
frequency space, we take advantage of a time formulation.
To obtain  a renormalized frequency scale $\bar\gamma$, we integrate the
length of a dipole from zero up to $1/\bar\gamma$, where the inverse
time scale $\bar\gamma$ is self-consistently given by the short-time part of
the breathing mode integral. We have
\begin{equation}
\bar\gamma = 2\pi\epsilon V_0^2 \int_0^{1/\bar\gamma} d\tau\,
\frac{1}{(\omega_{\rm c} \tau)^{1-2\epsilon}}
 = \frac{\pi V_0^2}{\omega_{\rm c}} \left(\frac{\omega_{\rm c}}{
\bar\gamma}\right)^{2\epsilon} \; .
\end{equation}
From this we obtain
\[
\bar\gamma = \gamma\,(\omega_{\rm c}/\gamma)^{2\epsilon}\;,\qquad
{\rm where}\;\quad \gamma = \pi V_0^2/\omega_{\rm c}\;.
\]
The frequency $\bar\gamma$ is the effective inverse time scale of the problem,
and $\gamma$ is the corresponding quantity when $\epsilon =0$.
The related temperature $\bar\gamma/k_{\rm B}$
coincides with the previously introduced Kondo temperature (\ref{kondo})
for $g=1/2-\epsilon$.

To develop a general computational strategy, it is convenient to split up
every time integral
over the length of a dipole into a contribution from the short-time regime
$0<\tau<1/\bar\gamma$ and a residual contribution from lengths
$\tau> 1/\bar\gamma$. We shall refer to the short-time part
as {\em collapsed dipole}. The crucial observation guiding
us towards this terminology is as follows.
 Within leading-log accuracy, a collapsed dipole
has zero dipole moment and therefore {\em no} interactions
with other charges. This property allows us to carry out the grand
canonical sum over all possible arrangements of collapsed
dipoles between two confining charges in an exact manner.
 The strategy is then completed by
developing a systematic scheme to calculate the contributions from all the
residual time intervals $\tau_j > 1/\bar\gamma$ in the asymptotic low-energy
regime (\ref{conditions}).

There are two types of collapsed dipoles, namely $(+-)$ and
$(-+)$. If they describe hops forth and back from states labelled by  an
{\em even} $p$ value, the contributions of these two dipoles  cancel
each other completely. This is
caused by the different sign of the phase factors $-\pi\epsilon p$ and
$\pi\epsilon p$ which are connected with the dipoles $(-+)$ and
$(+-)$, respectively. Hence there is no insertion of
a gas of collapsed dipoles for the {\em even} time
intervals $\tau_2,\tau_4,\ldots$
in Eq.~(\ref{udef}), where the path is in an even $p$ state.
However, if the dipoles describe hops forth and back from states
labelled by {\em odd}\, $p$, the dipoles $(-+)$ and $(+-)$ come with
phase factors $\pi\epsilon (1+p)$ and $\pi\epsilon(1-p)$,
respectively. Note that the second type does not contribute when
$p=1$. The sum of both contributions is independent of $p$. Allowing
next that the collapsed dipole $(-+)$
[as well as $(+-)$] moves freely within an
interval $\tau$, we find a total contribution $-\bar{\gamma}\tau$.
Finally, if we consider instead of the two dipoles  a grand canonical
noninteracting gas of these being confined within an odd-time interval $\tau$,
the total contribution is summed to an exponential factor
$\exp(-\bar\gamma \tau)$.

To summarize these findings, the overall effect of the collapsed dipoles is
to cause  exponential suppression of every long odd time interval on a length
$1/\bar{\gamma}$, while long even time intervals are {\em not}\,
suppressed on this scale.
In the diagrammatic expansion given below, we indicate the
insertion of a grand canonical gas of collapsed dipoles by a
square box. We shall refer to {\em bare} diagrams when they are free of
collapsed dipoles and to {\em dressed} diagrams when every odd time
interval is dressed by a square box.
\section{Diagrammatic expansion} \label{sec:3:1}
In the following, we wish to evaluate the exact expression 
(\ref{kinkgas}) with (\ref{udef})
for the nonlinear conductance in the low-energy regime specified
in Eq.~(\ref{conditions}). In that case, a leading-log summation
becomes possible around $g=1/2$. With
$g=1/2-\epsilon$, we compute the coefficients $a_j$ in
the series
\begin{equation}
\label{leadlog}
G(V,T) = \sum_{j=0}^\infty a_j(V,T)\,
 \epsilon^j_{} \ln^j[\mbox{max}(k_{\rm B}T,eV)] \;.
\end{equation}
Contributions $\sim \epsilon^k_{} \ln^j[\mbox{max}(k_{\rm B}T,
eV)]$ with $j<k$ are disregarded in our calculation.
This approximation is valid under conditions (\ref{conditions}).
In this section, computation of the functions $a_j(V,T)$ is
presented in detail.

\begin{figure}
\thicklines
\begin{picture}(7,5.5)
\put(0,3){\line(1,0){3.0}}
\put(4,3){\line(1,0){3.0}}
\put(1.5,1){\line(0,1){2.0}}
\put(5.5,3){\line(0,1){2.0}}
\put(3.5,3){\usebox{\cdg}}
\end{picture}
\caption[]{\label{fig2} Diagram giving the
contribution $U_1$ to the conductance. The square
stands for a factor $\exp[-\bar\gamma\tau]$
due to a gas of collapsed dipoles.
}
\end{figure}
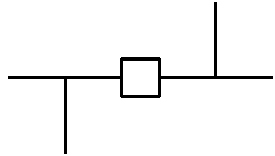

We start by rewriting the quantity $U(V,T)$ defined in
(\ref{udef}) in terms of a diagrammatic expansion,
\begin{equation}\label{diaexp}
U(V,T) = \sum_{n=1}^\infty U_n\; .
\end{equation}
The quantity $U_n$ is the sum of all arrangements of $2n$ charges
complying with the rules stated above.
The procedure is then as follows. (1) We have to draw all possible bare
diagrams with $2n$ charges of total charge zero which cannot be subdivided
into disjoint neutral parts, and the first charge has to be $\xi_1=-1$.
(2) In the next step, the bare diagrams are dressed
by decorating all odd time intervals $\tau_{2j-1}$ $(j=1,\ldots,n)$ 
with square boxes.
Each of the square boxes symbolizes insertion of a grand canonical gas of
collapsed dipoles, i.e., a factor of $\exp[-\bar\gamma \tau_{2j-1}]$.

\begin{figure}
\thicklines
\begin{picture}(13,6.5)
\put(0,4){\line(1,0){3.0}}
\put(4,4){\line(1,0){5.0}}
\put(10,4){\line(1,0){3.0}}
\put(1.5,2){\line(0,1){2.0}}
\put(5.5,2){\line(0,1){2.0}}
\put(7.5,4){\line(0,1){2.0}}
\put(11.5,4){\line(0,1){2.0}}
\put(3.5,4){\usebox{\cdg}}
\put(9.5,4){\usebox{\cdg}}
\end{picture}
\caption[]{\label{fig3} Diagram giving the
 contribution $U_2$.}
\end{figure}
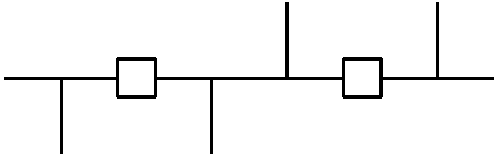

As outlined before, there are no such insertions for
even  time intervals $\tau_{2j}$. To avoid double counting,
integrations over even times are free of collapsed contributions,
and therefore we restrict these integrations to
$\bar\gamma\tau^{}_{2j} > 1$.
The term $U_n$  contributes to order $\epsilon^{n-1}$ (and to
higher orders).  To illustrate this, the diagrams representing
the terms $U_1$ to $U_4$ are shown in Figs.~\ref{fig2} to \ref{fig5}.
While in orders $n=1$ and $n=2$ only one diagram is found, respectively,
there are two diagrams for $n=3$ and five diagrams for $n=4$.

\begin{figure}
\thicklines
\begin{picture}(16,4)
\put(0.5,3.5){(I)}
\put(0,2){\line(1,0){2}}
\put(3,2){\line(1,0){4}}
\put(8,2){\line(1,0){4}}
\put(13,2){\line(1,0){2}}
\put(1,0){\line(0,1){2.0}}
\put(4,0){\line(0,1){2.0}}
\put(6,0){\line(0,1){2.0}}
\put(9,2){\line(0,1){2.0}}
\put(11,2){\line(0,1){2.0}}
\put(14,2){\line(0,1){2.0}}
\put(2.5,2){\usebox{\cdg}}
\put(7.5,2){\usebox{\cdg}}
\put(12.5,2){\usebox{\cdg}}
\end{picture}
\end{figure}
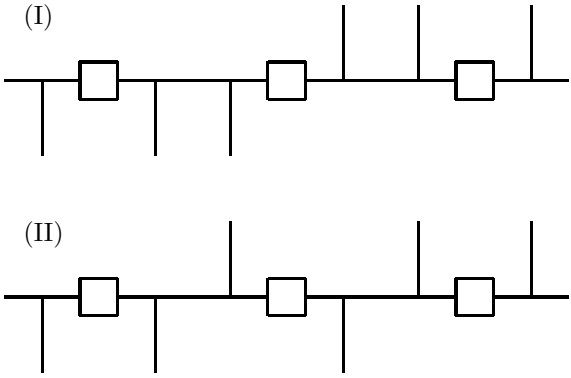
\begin{figure}
\thicklines
\begin{picture}(16,4.5)
\put(0.5,4){(II)}
\put(0,2.5){\line(1,0){2}}
\put(3,2.5){\line(1,0){4}}
\put(8,2.5){\line(1,0){4}}
\put(13,2.5){\line(1,0){2}}
\put(1,0.5){\line(0,1){2.0}}
\put(4,0.5){\line(0,1){2.0}}
\put(6,2.5){\line(0,1){2.0}}
\put(9,0.5){\line(0,1){2.0}}
\put(11,2.5){\line(0,1){2.0}}
\put(14,2.5){\line(0,1){2.0}}
\put(2.5,2.5){\usebox{\cdg}}
\put(7.5,2.5){\usebox{\cdg}}
\put(12.5,2.5){\usebox{\cdg}}
\end{picture}
\caption[]{\label{fig4} The 2 diagrams  yielding  the
contribution $U_3$.}
\end{figure}

The dressed diagram $U_1$ yields already the exact solution for $g=1/2$
since all other contributions are at least of order $\epsilon$.
From Eq.~(\ref{udef}), we find
\[
U_1=-V_0^2 \int_0^\infty
 d\tau \,\exp[-(\bar\gamma+ieV/2)\tau - S(\tau)]\;,
\]
which gives indeed the exact solution for $g=1/2$ in terms of the
digamma function $\psi(z)$,
\begin{equation}\label{exactsol}
G(V,T)/G_0 = 1 - \frac{2\gamma}{eV}
\,{\rm Im}\, \psi\left(\frac{1}{2} + \frac{\gamma+ieV/2}
{2\pi k_{\rm B} T} \right)\;.
\end{equation}
Thus the notion of collapsed dipoles provides a remarkably simple
derivation of this important result.\cite{weiss}
For finite $\epsilon$, the quantity $U_1$ is the same as
for $g=1/2$ within leading-log accuracy, except for $\gamma$ being
replaced by $\bar\gamma$. At this point, we would like to remark that
the diagram $U_1$ gives the correct
limiting behavior $G/G_0 \to 0$ as $T\to 0,\, V \to 0$
for any $V_0$.
Hence all higher-order dressed diagrams should
give vanishing contributions in the zero-energy limit, and this indeed
we find.

\begin{figure}
\thicklines
\begin{picture}(16,5)
\put(0.1,3){(I)}
\put(0,2){\line(1,0){1.5}}
\put(2.5,2){\line(1,0){3}}
\put(6.5,2){\line(1,0){3}}
\put(10.5,2){\line(1,0){3}}
\put(14.5,2){\line(1,0){1.5}}
\put(1,0){\line(0,1){2.0}}
\put(3,0){\line(0,1){2.0}}
\put(5,0){\line(0,1){2.0}}
\put(7,2){\line(0,1){2.0}}
\put(9,0){\line(0,1){2.0}}
\put(11,2){\line(0,1){2.0}}
\put(13,2){\line(0,1){2.0}}
\put(15,2){\line(0,1){2.0}}
\put(2,2){\usebox{\cdg}}
\put(6,2){\usebox{\cdg}}
\put(10,2){\usebox{\cdg}}
\put(14,2){\usebox{\cdg}}
\end{picture}
\end{figure}
\begin{figure}
\thicklines
\begin{picture}(16,5)
\put(0.1,3){(II)}
\put(0,2){\line(1,0){1.5}}
\put(2.5,2){\line(1,0){3}}
\put(6.5,2){\line(1,0){3}}
\put(10.5,2){\line(1,0){3}}
\put(14.5,2){\line(1,0){1.5}}
\put(1,0){\line(0,1){2.0}}
\put(3,0){\line(0,1){2.0}}
\put(5,2){\line(0,1){2.0}}
\put(7,0){\line(0,1){2.0}}
\put(9,2){\line(0,1){2.0}}
\put(11,0){\line(0,1){2.0}}
\put(13,2){\line(0,1){2.0}}
\put(15,2){\line(0,1){2.0}}
\put(2,2){\usebox{\cdg}}
\put(6,2){\usebox{\cdg}}
\put(10,2){\usebox{\cdg}}
\put(14,2){\usebox{\cdg}}
\end{picture}
\end{figure}
\begin{figure}
\thicklines
\begin{picture}(16,5)
\put(0.1,3){(III)}
\put(0,2){\line(1,0){1.5}}
\put(2.5,2){\line(1,0){3}}
\put(6.5,2){\line(1,0){3}}
\put(10.5,2){\line(1,0){3}}
\put(14.5,2){\line(1,0){1.5}}
\put(1,0){\line(0,1){2.0}}
\put(3,0){\line(0,1){2.0}}
\put(5,0){\line(0,1){2.0}}
\put(7,2){\line(0,1){2.0}}
\put(9,2){\line(0,1){2.0}}
\put(11,0){\line(0,1){2.0}}
\put(13,2){\line(0,1){2.0}}
\put(15,2){\line(0,1){2.0}}
\put(2,2){\usebox{\cdg}}
\put(6,2){\usebox{\cdg}}
\put(10,2){\usebox{\cdg}}
\put(14,2){\usebox{\cdg}}
\end{picture}
\end{figure}
\begin{figure}
\thicklines
\begin{picture}(16,6)
\put(0.1,4){(IV)}
\put(0,3){\line(1,0){1.5}}
\put(2.5,3){\line(1,0){3}}
\put(6.5,3){\line(1,0){3}}
\put(10.5,3){\line(1,0){3}}
\put(14.5,3){\line(1,0){1.5}}
\put(1,1){\line(0,1){2.0}}
\put(3,1){\line(0,1){2.0}}
\put(5,3){\line(0,1){2.0}}
\put(7,1){\line(0,1){2.0}}
\put(9,1){\line(0,1){2.0}}
\put(11,3){\line(0,1){2.0}}
\put(13,3){\line(0,1){2.0}}
\put(15,3){\line(0,1){2.0}}
\put(2,3){\usebox{\cdg}}
\put(6,3){\usebox{\cdg}}
\put(10,3){\usebox{\cdg}}
\put(14,3){\usebox{\cdg}}
\end{picture}
\end{figure}
\begin{figure}
\thicklines
\begin{picture}(16,6)
\put(0.1,4){(V)}
\put(0,3){\line(1,0){1.5}}
\put(2.5,3){\line(1,0){3}}
\put(6.5,3){\line(1,0){3}}
\put(10.5,3){\line(1,0){3}}
\put(14.5,3){\line(1,0){1.5}}
\put(1,1){\line(0,1){2.0}}
\put(3,1){\line(0,1){2.0}}
\put(5,1){\line(0,1){2.0}}
\put(7,1){\line(0,1){2.0}}
\put(9,3){\line(0,1){2.0}}
\put(11,3){\line(0,1){2.0}}
\put(13,3){\line(0,1){2.0}}
\put(15,3){\line(0,1){2.0}}
\put(2,3){\usebox{\cdg}}
\put(6,3){\usebox{\cdg}}
\put(10,3){\usebox{\cdg}}
\put(14,3){\usebox{\cdg}}
\end{picture}
\caption[]{\label{fig5}
The 5 diagrams contributing to $U_4$. Diagrams (I) -- (IV) are of
type (A), and diagram (V) is of type (B).}
\end{figure}

Generally, besides the lowest-order diagram in Fig.~\ref{fig1}
we may distinguish two types of dressed diagrams.
Diagrams of type (A) are of the form $(--X++)$. Here, $X$ stands for
insertion of any arrangement of {\em extended dipoles}
$(+-)$ or $(-+)$ between the outer
double charges $(--)$ and $(++)$.\cite{footn}
 Since there are two types of extended
dipoles, we have $2^{n-2}$ dressed diagrams of type
(A) contributing to order $U_n$. These dipoles are of finite length in
contrast to the collapsed dipoles considered before. Hence they are
interacting with each other and also with the outer charge pairs. However,
because of the insertions of collapsed dipoles (indicated by the squares
in the diagrams), they are narrow compared to typical inter-dipole
distances. All other diagrams contain at least four pairs of equal
charge. We shall refer to them as diagrams of type (B). The simplest
one of this type contributes to $U_4$ and is diagram
(V) in Fig.~\ref{fig5}. Remarkably, up to order $U_3$ only diagrams
of type (A) are encountered.

Now we have to extract and evaluate the leading logarithmic
contributions in the terms with $n > 1$.
It simplifies notation to use scaled times and energies henceforth,
\begin{equation}\label{scales}
x_j= \bar\gamma\tau_j\;,\quad u =
2\pi k_{\rm B} T/\bar\gamma
\;,\quad  v = 2igeV /\bar\gamma\;.
\end{equation}
According to the conditions (\ref{conditions}), we are interested in the
asymptotic regime
\begin{equation}\label{cond}
u \ll 1\; ; \quad\qquad |v| \ll 1 \; .
\end{equation}
Now every {\em odd}\, time integration
 $\int dx_{2j+1}^{}$ has a weight function
$\exp(-x_{2j+1}^{})$ due to the insertion of collapsed dipoles while all
{\em even} time integrations are free of this weight. Hence it
follows that the expression $U_n$ is dominated by the regime
\begin{equation}\label{regime}
 x_{2j+1}^{} \ll x_{2k}^{} \; , \qquad\quad j,k=0,1,\ldots, n-1\;,
\end{equation}
and the logarithms we wish to extract must arise
from the integrations over the long even times.
In the following, we describe in detail the general strategy we
have employed to evaluate diagrams like the ones shown in
Figs.~\ref{fig2}--\ref{fig5}.

We first rewrite the interaction factor (\ref{intfac})
using the definitions
\begin{eqnarray}
M_n &=& \exp\sum_{j>k=1}^{2n}
             \xi_j R(x_{jk})\xi_k \; , \label{mm}\\
N_n &=& (2/u)^{2n\epsilon} M_n^{-2\epsilon} \; ,
\label{nm} \\
R(x) &=& \sinh(ux/2) \; , \label{rdef}
\end{eqnarray}
such that
\begin{equation} \label{intf}
W_n = (\bar\gamma/\omega_{\rm c})^{(1-2\epsilon)n}\, (u/2)^n \,M_n N_n \; .
\end{equation}
The factor $M_n$ includes the interactions of $2n$ charges
for $g=1/2$, while the factor $N_n$ encapsulates the remaining
$\epsilon$-dependent part of the interactions. The usefulness
of the split-up (\ref{intf}) is demonstrated now.

Having drawn a certain diagram contributing to $U_n$,
say $U_n^{(j)}$, we take advantage
of a general decomposition theorem for the interaction factor
$M_n$ in (\ref{mm}). The decomposition theorem corresponds to
Wick$^\prime$s theorem\cite{fetter} for fermionic fields and originates
 from the equivalence of the interaction factor $\exp(-S(\tau))$
for $g=1/2$ with a free 1D fermion propagator.\cite{leggett87}
 By virtue of this theorem, we can decompose
$M_n$ --- which contains $n(2n-1)$ pair interactions --- into a
sum of $n!$ terms containing only $n$ intrapair interactions of
neutral pairs. This procedure results in the corresponding
decomposition of a given $n$th order diagram into $n!$ graphs.
The split-up (\ref{intf}) allows us therefore to {\em topologically
reduce the diagrams to graphs which are much easier to evaluate}.
Thereby the quantity $U_n$ takes the form of a sum over all diagrams
(labelled by an index $(j)$) where each diagram is represented by a sum
of $n!$ graphs,
\begin{equation} \label{graphs}
U_n= \frac{\epsilon^{n-1} u^n \bar\gamma}{2\pi}\;
\sum_j\sum_{r=1}^{n!} U^{(j,r)}_n\;,
\end{equation}
where $U^{(j,r)}_n$ describes the contribution of the $r$th graph
to the $n$th-order diagram $j$ under study (with the prefactor
chosen accordingly).
Pictorially, one obtains the graphs by grouping the $2n$ charges into
$n$ neutral pairs. Clearly, there are $n!$ different assignments and
hence $n!$ different graphs for each diagram,
with the sign of each graph given by the
number of crossings of lines connecting the paired charges.
Fig.~\ref{fig6} shows the 2 graphs for the diagram in Fig.~\ref{fig3}
resulting in $U_2$, and in Fig.~\ref{fig7} the 12 graphs
due to the two diagrams of $U_3$ are displayed.

In the final step, we group all graphs $U_n^{(j,r)}$ into {\em classes} with
respect to their particular dependence on the even time intervals
 $x_{2j}^{}$. Each of these classes is then evaluated in leading-log
accuracy, and the remaining odd time integrations can be carried out
without further approximation.

So far we have not taken into account the residual interaction factor
$N_n$ given in (\ref{nm}). We have to treat this term separately
as it does not fit into the decomposition scheme. Under
condition (\ref{regime}) and within leading-log accuracy, this factor has
the same form for all diagrams of type (A) and is
\begin{equation}\label{nnl}
N_n^{(A)} = (x_2^{} + x_4^{} +\cdots + x_{2n-2}^{})^{8 \epsilon}\; .
\end{equation}
Unfortunately, for graphs of type (B), it is not so straightforward to
include the factor $N_n$ in the explicit calculation. Therefore,
in this paper, we confine
ourselves to the regime where it is consistent to
put $N^{(B)}_n =1$ (see Sec.~\ref{sec:4} and Appendix).

\section{LEADING-LOG EVALUATION OF THE CONDUCTANCE}

Following the general diagrammatic approach outlined in the
previous section, we have explicitly calculated the leading logs in
the expressions $U_n$ for $n\leq 4$. For larger $n$,
the explicit calculation of all diagrams
becomes prohibitively cumbersome. However, for the asymptotic
low-energy corrections only the restricted set of diagrams of type (A)
contributes, as we shall explain below. This permits an exact
evaluation of the asymptotic corrections.
The corresponding summation of all diagrams of type (A) is carried out
in Sec.~\ref{sec:a}. In the remainder
of this section, we illustrate the computation of $U_n$
for $n\leq 4$. That will also lead us towards the
exact solution of this problem, see Sec.~\ref{sec:b}.

\subsection{Contribution $U_2$}\label{sec:22}

To proceed according to the strategy of Sec.~III, we have to
first draw all diagrams in order $n=2$. There is only the
single diagram shown in Fig.~\ref{fig3}, which is of type (A)
and contributes in order $\epsilon$ (and higher orders) to
the series (\ref{leadlog}). In a second step, this diagram
may be topologically reduced into the two graphs
shown in Fig.~\ref{fig6}, as follows from the decomposition of the
interaction factor
\begin{eqnarray}
M_2 &=& e_{}^{R(x_1^{})+R(x_3^{})-R(x_2^{})-R(x_1^{}+x_2^{}) -
R(x_2^{} +x_3^{}) - R(x_1^{} +x_2^{} + x_3^{}) } \nonumber\\
&= & e_{}^{-R(x_2^{}) - R(x_1^{}  + x_2^{} + x_3^{}) }
 - e_{}^{- R(x_1^{} + x_2^{}) - R(x_2^{} + x_3^{}) } \; , \nonumber
\end{eqnarray}
where $R(x)$ is given by Eq.~(\ref{rdef}).
Both graphs fall into different classes, i.e.~they have a
different functional dependence on the long even time interval $x_2$.
In particular, according to Eqs.~(\ref{udef}) and (\ref{graphs}),
 we have to evaluate the terms
\begin{eqnarray*}
U^{(1)}_2&=& \int_0^\infty dx_1 dx_3 \int_1^\infty
dx_2 \\&\times&  \frac{ x_2^{8\epsilon}\, e^{-[1+v/2](x_1+x_3) -
vx_2}}{\sinh[u(x_2+x_1+x_3)/2] \sinh[ux_2/2]} \; , \\
U^{(2)}_2&=& - \int_0^\infty dx_1 dx_3 \int_1^\infty  dx_2
 \\ &\times& \frac{x_2^{8\epsilon}\,e^{-[1+v/2](x_1+x_3) -
vx_2}}{\sinh[u(x_2+x_1)/2] \sinh[u(x_2+x_3)/2]} \;,
\end{eqnarray*}
where we have used (\ref{nnl}).
Recalling that the dominant contributions to the triple integrals come
from the region $x_1,x_3\ll x_2$, it is clear that the logarithms which
we wish to extract arise from the $x_2$-integrals. To proceed, we use
the approximation
\begin{eqnarray} \label{helpf} &&
\frac{u \exp[ux+u(a+b)/2]}{4\sinh[u(x+a)/2]
\sinh[u(x+b)/2]} \approx \\ \nonumber && \quad
\frac{(1-e^{-u(b-a)})^{-1}}{x+a} +
\frac{(1-e^{u(b-a)})^{-1}}{x+b} \;,
\end{eqnarray}
where $x$ corresponds to the long even time $x_2$
and $a,b$ are linear combinations of short odd times $x_1,x_3$.
The terms neglected on the r.h.s.~are subleading as they do not
give logarithmic contributions. It is at this point where we
invoke the leading-logarithmic approximation.

Identifying $a$ and $b$ with $q$, the contributions to the integral
over the even time are in the form
\begin{equation}\label{even}
\int_1^\infty dx \frac{x^{8\epsilon}}{x+q} e^{-(u+v)x} =
e^{q(u+v)} L_2(\epsilon)\; .
\end{equation}
Since the odd time integrations
produce a purely imaginary contribution, we need to consider only
the real part of $L_2(\epsilon)$, i.e.~we may replace
$u+v$ by $|u+v|$ in this term when performing these integrations.
Within leading-log accuracy, we have
\begin{equation} \label{log1}
L_2(\epsilon) = \int_1^\infty dx \frac{e^{-|u+v|x}}{x^{1-8\epsilon}}
= -\ln|u+v|\,\Bigl( 1 + O(\epsilon\ln|u+v|)\Bigr) \;.
\end{equation}

\begin{figure}
\thicklines
\begin{picture}(13,7.5)
\put(0.1,7){(1)}
\put(0,4){\line(1,0){3.0}}
\put(4,4){\line(1,0){5.0}}
\put(10,4){\line(1,0){3.0}}
\put(1.5,2){\line(0,1){2.0}}
\put(5.5,2){\line(0,1){2.0}}
\put(7.5,4){\line(0,1){2.0}}
\put(11.5,4){\line(0,1){2.0}}
\put(6.5,6.2){\oval(1.9,1.2)[t]}
\put(6.5,6.6){\oval(10,1.1)[t]}
\put(3.5,4){\usebox{\cdg}}
\put(9.5,4){\usebox{\cdg}}
\end{picture}
\end{figure}
\begin{figure}
\thicklines
\begin{picture}(13,7.5)
\put(0.1,7){(2)}
\put(0,4){\line(1,0){3.0}}
\put(4,4){\line(1,0){5.0}}
\put(10,4){\line(1,0){3.0}}
\put(1.5,2){\line(0,1){2.0}}
\put(5.5,2){\line(0,1){2.0}}
\put(7.5,4){\line(0,1){2.0}}
\put(11.5,4){\line(0,1){2.0}}
\put(4.5,6.6){\oval(6,1.2)[t]}
\put(8.5,6.4){\oval(5.9,1.2)[t]}
\put(3.5,4){\usebox{\cdg}}
\put(9.5,4){\usebox{\cdg}}
\end{picture}
\caption[]{\label{fig6} The two graphs obtained from the decomposition
theorem for the contribution $U_2$ shown in Fig.~\ref{fig3}. The
relative signs are $+$ for graph (1) and $-$ for graph (2).
The curves display interactions between paired charges.  }
\end{figure}
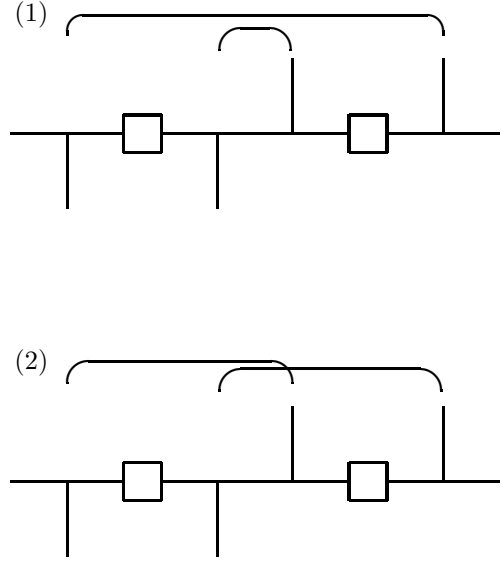

Upon using Eqs.~(\ref{helpf}) and (\ref{even}), the resulting expressions
for $U^{(1)}_2$ and $U^{(2)}_2$ may be combined in a form
which will find straightforward generalization to higher orders,
\begin{eqnarray}\label{a2}
{\rm Im} \sum_r U^{(r)}_2
 &=& -\frac{2}{u} L_2(\epsilon) \int_0^\infty dx_1 dx_3
\,e^{-(x_1 +x_3)}_{}\\
 &\times& \sum_{\{\sigma_j =\pm 1\}} \sigma_1 \sigma_3
\frac{ \sin[|v|(\sigma_1x_1 + \sigma_3x_3)/2]}{
          \sinh[u(\sigma_1x_1 + \sigma_3x_3)/2]} \; . \nonumber
\end{eqnarray}
The remaining odd time integrations pose no problem and are done
easily.

Collecting all terms together, we find the leading-log contributions to
the conductance resulting from $U_2$ in all orders in $\epsilon$,
\begin{eqnarray} \nonumber
{\rm Im}\,U_2 &=&
4\epsilon\, \frac{\bar\gamma}{\pi} \, L_2(\epsilon) \,
{\rm Im}\Bigl[ \psi\left(\frac{2+u+v}{2u}\right)\\ &&\quad  +u^{-1} \psi'
\left(\frac{2+u+v}{2u}\right)\Bigr] \;.  \label{u2}
\end{eqnarray}
The function $\psi'$ is the derivative of the $\psi$ function.
Together with the contribution due to  $U_1$, this
gives {\em all} terms in order $\epsilon$ (plus partial summation of
higher-order terms in $\epsilon$).
\subsection{Contribution $U_3$} \label{sec:33}
To proceed further, let us now study the next order. The
two diagrams contributing to order $U_3$ are
displayed in Fig.~\ref{fig4}. Applying the before-mentioned decomposition
scheme for the interaction factor $M_3$, we find the
12 graphs shown in Fig.~\ref{fig7}.
Again we group them into classes regarding their
dependence on the even intervals $x_2, x_4$.
The contributions $U^{(j,r)}_3 \;(j=1,2;\,r=1,\ldots,6)$
corresponding to the 12 graphs in Fig.~\ref{fig7}
are expressed as
\begin{eqnarray*}
U^{(1,\,r=1,\ldots,6)}_3 &=& \int_0^\infty dx_1 dx_3 dx_5 \int_1^\infty
dx_2 dx_4 \\ &\times&e^{-[1+v/2](x_1+x_3+x_5)-vx_3} \,
e^{- v(x_2+x_4)} \,N^{(A)}_{3}  f_3^{(1,r)} \\
U^{(2,\,r=1,\ldots,6)}_3 &=& \int_0^\infty dx_1 dx_3 dx_5 \int_1^\infty
dx_2 dx_4 \\ &\times& e^{-[1+v/2](x_1+x_3+x_5)} \,
e^{- v(x_2+x_4)} \,N^{(A)}_{3}  f_3^{(2,r)} \; ,
\end{eqnarray*}
where $N^{(A)}_n$ is given in (\ref{nnl}), and where
each $f_3^{(j,r)}$ belongs to one of the following two classes,
\begin{eqnarray*}
f_3^{(1)}& =&
 \Bigl( \sinh[u(x_2+x_4+a)/2] \\
&\times& \sinh[u(x_2+x_4+b)/2]\sinh[uc/2]\Bigr)^{-1}\;,\\
f_3^{(2)}& =&
 \Bigl( \sinh[u(x_2+a)/2] \\ &\times&
\sinh[u(x_4+b)/2]\sinh[u(x_2+x_4+c)/2]\Bigr)^{-1}\;.
\end{eqnarray*}
Here, $a,b,c$ denote certain linear combinations of the short
odd time intervals $x_1,x_3,x_5\ll x_2,x_4$. This classification
is  again motivated by the fact that logarithmic terms are
attributed to the domain of long even time intervals.

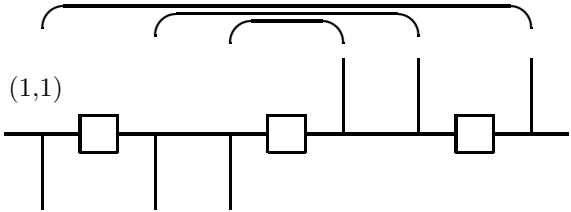
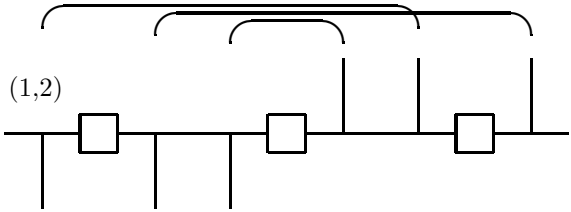
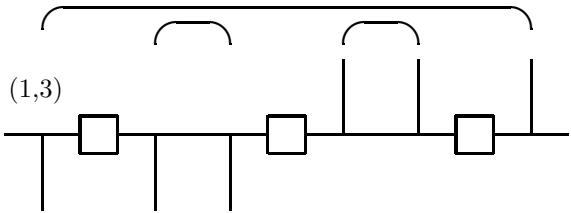
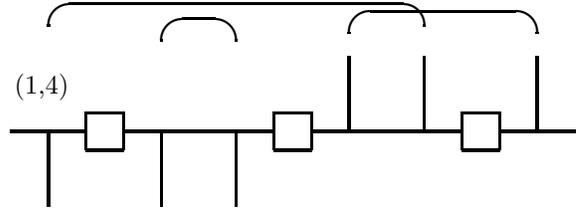
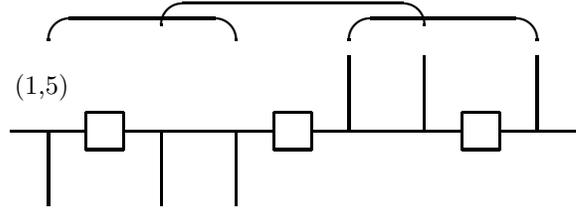
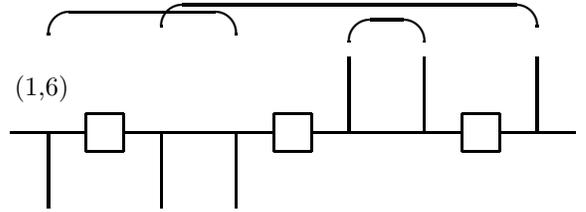
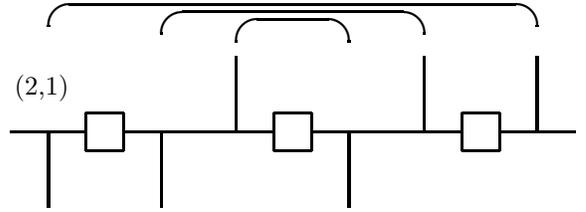
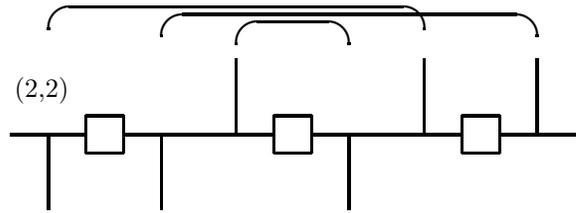
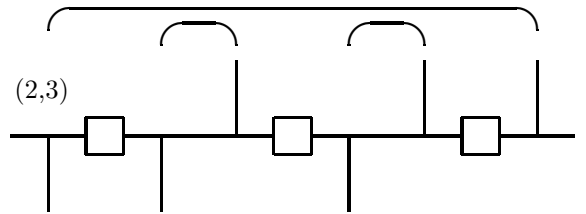
\begin{figure}
\thicklines
\begin{picture}(16,6)
\put(0.1,3){(1,1)}
\put(0,2){\line(1,0){2}}
\put(3,2){\line(1,0){4}}
\put(8,2){\line(1,0){4}}
\put(13,2){\line(1,0){2}}
\put(1,0){\line(0,1){2.0}}
\put(4,0){\line(0,1){2.0}}
\put(6,0){\line(0,1){2.0}}
\put(9,2){\line(0,1){2.0}}
\put(11,2){\line(0,1){2.0}}
\put(14,2){\line(0,1){2.0}}
\put(2.5,2){\usebox{\cdg}}
\put(7.5,2){\usebox{\cdg}}
\put(12.5,2){\usebox{\cdg}}
\put(7.5,4.8){\oval(13,1.2)[t]}
\put(7.5,4.6){\oval(7,1.2)[t]}
\put(7.5,4.4){\oval(3,1.2)[t]}
\end{picture}
\end{figure}
\begin{figure}
\thicklines
\begin{picture}(16,6)
\put(0.1,3){(1,2)}
\put(0,2){\line(1,0){2}}
\put(3,2){\line(1,0){4}}
\put(8,2){\line(1,0){4}}
\put(13,2){\line(1,0){2}}
\put(1,0){\line(0,1){2.0}}
\put(4,0){\line(0,1){2.0}}
\put(6,0){\line(0,1){2.0}}
\put(9,2){\line(0,1){2.0}}
\put(11,2){\line(0,1){2.0}}
\put(14,2){\line(0,1){2.0}}
\put(2.5,2){\usebox{\cdg}}
\put(7.5,2){\usebox{\cdg}}
\put(12.5,2){\usebox{\cdg}}
\put(6,4.8){\oval(10,1.2)[t]}
\put(9,4.6){\oval(10,1.2)[t]}
\put(7.5,4.4){\oval(3,1.2)[t]}
\end{picture}
\end{figure}
\begin{figure}
\thicklines
\begin{picture}(16,6)
\put(0.1,3){(1,3)}
\put(0,2){\line(1,0){2}}
\put(3,2){\line(1,0){4}}
\put(8,2){\line(1,0){4}}
\put(13,2){\line(1,0){2}}
\put(1,0){\line(0,1){2.0}}
\put(4,0){\line(0,1){2.0}}
\put(6,0){\line(0,1){2.0}}
\put(9,2){\line(0,1){2.0}}
\put(11,2){\line(0,1){2.0}}
\put(14,2){\line(0,1){2.0}}
\put(2.5,2){\usebox{\cdg}}
\put(7.5,2){\usebox{\cdg}}
\put(12.5,2){\usebox{\cdg}}
\put(7.5,4.8){\oval(13,1.2)[t]}
\put(10,4.4){\oval(2,1.2)[t]}
\put(5,4.4){\oval(2,1.2)[t]}
\end{picture}
\end{figure}
\begin{figure}
\thicklines
\begin{picture}(16,6)
\put(0.1,3){(1,4)}
\put(0,2){\line(1,0){2}}
\put(3,2){\line(1,0){4}}
\put(8,2){\line(1,0){4}}
\put(13,2){\line(1,0){2}}
\put(1,0){\line(0,1){2.0}}
\put(4,0){\line(0,1){2.0}}
\put(6,0){\line(0,1){2.0}}
\put(9,2){\line(0,1){2.0}}
\put(11,2){\line(0,1){2.0}}
\put(14,2){\line(0,1){2.0}}
\put(2.5,2){\usebox{\cdg}}
\put(7.5,2){\usebox{\cdg}}
\put(12.5,2){\usebox{\cdg}}
\put(6,4.8){\oval(10,1.2)[t]}
\put(11.5,4.6){\oval(5,1.2)[t]}
\put(5,4.4){\oval(2,1.2)[t]}
\end{picture}
\end{figure}
\begin{figure}
\thicklines
\begin{picture}(16,6)
\put(0.1,3){(1,5)}
\put(0,2){\line(1,0){2}}
\put(3,2){\line(1,0){4}}
\put(8,2){\line(1,0){4}}
\put(13,2){\line(1,0){2}}
\put(1,0){\line(0,1){2.0}}
\put(4,0){\line(0,1){2.0}}
\put(6,0){\line(0,1){2.0}}
\put(9,2){\line(0,1){2.0}}
\put(11,2){\line(0,1){2.0}}
\put(14,2){\line(0,1){2.0}}
\put(2.5,2){\usebox{\cdg}}
\put(7.5,2){\usebox{\cdg}}
\put(12.5,2){\usebox{\cdg}}
\put(7.5,4.8){\oval(7,1.2)[t]}
\put(11.5,4.4){\oval(5,1.2)[t]}
\put(3.5,4.4){\oval(5,1.2)[t]}
\end{picture}
\end{figure}
\begin{figure}
\thicklines
\begin{picture}(16,6)
\put(0.1,3){(1,6)}
\put(0,2){\line(1,0){2}}
\put(3,2){\line(1,0){4}}
\put(8,2){\line(1,0){4}}
\put(13,2){\line(1,0){2}}
\put(1,0){\line(0,1){2.0}}
\put(4,0){\line(0,1){2.0}}
\put(6,0){\line(0,1){2.0}}
\put(9,2){\line(0,1){2.0}}
\put(11,2){\line(0,1){2.0}}
\put(14,2){\line(0,1){2.0}}
\put(2.5,2){\usebox{\cdg}}
\put(7.5,2){\usebox{\cdg}}
\put(12.5,2){\usebox{\cdg}}
\put(9,4.8){\oval(10,1.2)[t]}
\put(10,4.4){\oval(2,1.2)[t]}
\put(3.5,4.6){\oval(5,1.2)[t]}
\end{picture}
\end{figure}
\begin{figure}
\thicklines
\begin{picture}(16,6)
\put(0.1,3){(2,1)}
\put(0,2){\line(1,0){2}}
\put(3,2){\line(1,0){4}}
\put(8,2){\line(1,0){4}}
\put(13,2){\line(1,0){2}}
\put(1,0){\line(0,1){2.0}}
\put(4,0){\line(0,1){2.0}}
\put(6,2){\line(0,1){2.0}}
\put(9,0){\line(0,1){2.0}}
\put(11,2){\line(0,1){2.0}}
\put(14,2){\line(0,1){2.0}}
\put(2.5,2){\usebox{\cdg}}
\put(7.5,2){\usebox{\cdg}}
\put(12.5,2){\usebox{\cdg}}
\put(7.5,4.8){\oval(13,1.2)[t]}
\put(7.5,4.6){\oval(7,1.2)[t]}
\put(7.5,4.4){\oval(3,1.2)[t]}
\end{picture}
\end{figure}
\begin{figure}
\thicklines
\begin{picture}(16,6)
\put(0.1,3){(2,2)}
\put(0,2){\line(1,0){2}}
\put(3,2){\line(1,0){4}}
\put(8,2){\line(1,0){4}}
\put(13,2){\line(1,0){2}}
\put(1,0){\line(0,1){2.0}}
\put(4,0){\line(0,1){2.0}}
\put(6,2){\line(0,1){2.0}}
\put(9,0){\line(0,1){2.0}}
\put(11,2){\line(0,1){2.0}}
\put(14,2){\line(0,1){2.0}}
\put(2.5,2){\usebox{\cdg}}
\put(7.5,2){\usebox{\cdg}}
\put(12.5,2){\usebox{\cdg}}
\put(6,4.8){\oval(10,1.2)[t]}
\put(9,4.6){\oval(10,1.2)[t]}
\put(7.5,4.4){\oval(3,1.2)[t]}
\end{picture}
\end{figure}
\begin{figure}
\thicklines
\begin{picture}(16,6)
\put(0.1,3){(2,3)}
\put(0,2){\line(1,0){2}}
\put(3,2){\line(1,0){4}}
\put(8,2){\line(1,0){4}}
\put(13,2){\line(1,0){2}}
\put(1,0){\line(0,1){2.0}}
\put(4,0){\line(0,1){2.0}}
\put(6,2){\line(0,1){2.0}}
\put(9,0){\line(0,1){2.0}}
\put(11,2){\line(0,1){2.0}}
\put(14,2){\line(0,1){2.0}}
\put(2.5,2){\usebox{\cdg}}
\put(7.5,2){\usebox{\cdg}}
\put(12.5,2){\usebox{\cdg}}
\put(7.5,4.8){\oval(13,1.2)[t]}
\put(10,4.4){\oval(2,1.2)[t]}
\put(5,4.4){\oval(2,1.2)[t]}
\end{picture}
\end{figure}
\begin{figure}
\thicklines
\begin{picture}(16,6)
\put(0.1,3){(2,4)}
\put(0,2){\line(1,0){2}}
\put(3,2){\line(1,0){4}}
\put(8,2){\line(1,0){4}}
\put(13,2){\line(1,0){2}}
\put(1,0){\line(0,1){2.0}}
\put(4,0){\line(0,1){2.0}}
\put(6,2){\line(0,1){2.0}}
\put(9,0){\line(0,1){2.0}}
\put(11,2){\line(0,1){2.0}}
\put(14,2){\line(0,1){2.0}}
\put(2.5,2){\usebox{\cdg}}
\put(7.5,2){\usebox{\cdg}}
\put(12.5,2){\usebox{\cdg}}
\put(6,4.8){\oval(10,1.2)[t]}
\put(11.5,4.6){\oval(5,1.2)[t]}
\put(5,4.4){\oval(2,1.2)[t]}
\end{picture}
\end{figure}
\begin{figure}
\thicklines
\begin{picture}(16,6)
\put(0.1,3){(2,5)}
\put(0,2){\line(1,0){2}}
\put(3,2){\line(1,0){4}}
\put(8,2){\line(1,0){4}}
\put(13,2){\line(1,0){2}}
\put(1,0){\line(0,1){2.0}}
\put(4,0){\line(0,1){2.0}}
\put(6,2){\line(0,1){2.0}}
\put(9,0){\line(0,1){2.0}}
\put(11,2){\line(0,1){2.0}}
\put(14,2){\line(0,1){2.0}}
\put(2.5,2){\usebox{\cdg}}
\put(7.5,2){\usebox{\cdg}}
\put(12.5,2){\usebox{\cdg}}
\put(7.5,4.8){\oval(7,1.2)[t]}
\put(11.5,4.4){\oval(5,1.2)[t]}
\put(3.5,4.4){\oval(5,1.2)[t]}
\end{picture}
\end{figure}
\begin{figure}
\thicklines
\begin{picture}(16,7)
\put(0.1,4){(2,6)}
\put(0,3){\line(1,0){2}}
\put(3,3){\line(1,0){4}}
\put(8,3){\line(1,0){4}}
\put(13,3){\line(1,0){2}}
\put(1,1){\line(0,1){2.0}}
\put(4,1){\line(0,1){2.0}}
\put(6,3){\line(0,1){2.0}}
\put(9,1){\line(0,1){2.0}}
\put(11,3){\line(0,1){2.0}}
\put(14,3){\line(0,1){2.0}}
\put(2.5,3){\usebox{\cdg}}
\put(7.5,3){\usebox{\cdg}}
\put(12.5,3){\usebox{\cdg}}
\put(9,5.8){\oval(10,1.2)[t]}
\put(10,5.4){\oval(2,1.2)[t]}
\put(3.5,5.6){\oval(5,1.2)[t]}
\end{picture}
\caption[]{\label{fig7}
The 12 graphs obtained from the 2 diagrams in order
$U_3$ shown in Fig.~\ref{fig4}. The graphs
$(1,1),(1,3),(1,5),(2,1),(2,3)$, and $(2,5)$ have positive sign,
the remaining ones are negative.}
\end{figure}
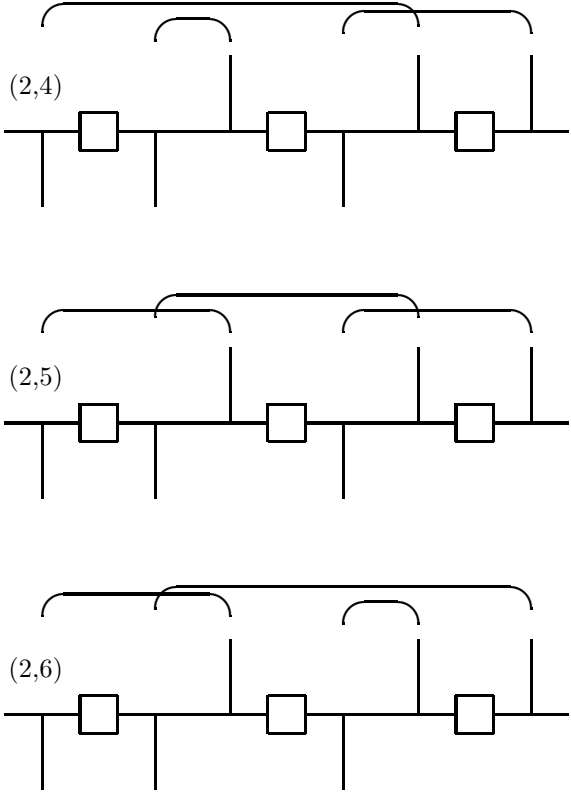

From Fig.~\ref{fig7} it is obvious that the graphs (1,1),(1,2),(2,1),
and (2,2) fall into the class $f_3^{(1)}$, whereas all other graphs belong to
the class $f_3^{(2)}$. To identify the classes giving leading logs, we put
$N^{(3)}_{A} =1$ for the moment. Then, in performing the even time
integrations, we have to look for terms giving squares of logarithms,
$ \ln^2|u+v|$. Upon using Eq.~(\ref{helpf}), it is easily shown that the
first class $f_3^{(1)}$ gives subleading terms  $\sim \ln|u+v|$ only, and
solely the eight diagrams in the second class produce leading logs.
The even time integrations are tackled by switching to the variables
\[
x=x_2+x_4 \;,\quad \rho = (x_2-x_4)/2\;,
\]
which allows for exact evaluation of the $\rho$ integral.
As the logarithms come from the region $x_1,x_3,x_5 \ll x$, the
$x$-integral takes the same form for the eight remaining graphs. Upon
performing a substitution $x+q \to x$ as given in (\ref{even}) with
(\ref{log1}), where now $q$ is $a$, $b$ or $c$, the logarithms of $|u+v|$ are
in the integral
\begin{eqnarray}\label{log2}
L_3(\epsilon)& =& \int_1^{\infty} dx\, \frac{\ln x}{x^{1-8\epsilon}}
\,e^{-|u+v|x} \\ \nonumber
 &=& \frac{1}{2} \ln^2 |u+v|\,\Bigl( 1 + O(\epsilon\ln|u+v|)\Bigr)\;,
\end{eqnarray}
where we have reintroduced $N^{(3)}_{A} = x^{8\epsilon}$.
Labelling the relevant 8 graphs with variables
$\sigma_1,\sigma_3,\sigma_5=\pm 1$, the imaginary part of the sum
of them can be written in the compact form
\begin{eqnarray}\nonumber 
&& {\rm Im}\, \sum_{j,r} U^{(j,r)}_3 =  \frac{16}{u^2} L_3(\epsilon)
\int_{0}^\infty dx_1 dx_3 dx_5\, e^{-(x_1+x_3+x_5)}\\&\times&
\label{kkk} \sum_{\{\sigma_j =\pm1\}}\sigma_1 \sigma_5\,
\frac{\sin[|v|(\sigma_1 x_1
+ \sigma_3 x_3 + \sigma_5 x_5)/2]}{\sinh[u(\sigma_1 x_1
+ \sigma_3 x_3 + \sigma_5 x_5)/2]} \;.
\end{eqnarray}
Here, the factor $\sigma_1  \sigma_5$ gives the signs
of the respective graphs.

To evaluate the remaining odd time integrations, it is useful to insert
the identity in form of
\begin{equation}\label{identity}
1 = \int_{-\infty}^\infty ds \int_{-\infty}^\infty \frac{dk}{2\pi}
\,e^{ik[s-(\sigma_1 x_1+\sigma_3 x_3 + \sigma_5 x_5)]}\;.
\end{equation}
 Performing the summation over the $\sigma^\prime$s and the
integration over $x_1,x_3,x_5$, one finds
a factor $-8k^2/[1+k^2]^3$, and we are left with
the $k$- and $s$-integrals. These can be carried out
analytically, and we obtain the leading-log terms of
$U_3$ in the form
\begin{eqnarray}\label{u33}
{\rm Im}\,U_3 &=&
-4\epsilon^2 \,\frac{\bar\gamma}{\pi}\,
L_3(\epsilon)\, {\rm Im}\Bigl[ \psi\left(\frac{2+u+v}{2u}\right) \\ &-&
  -u^{-1} \psi'\left(\frac{2+u+v}{2u}\right) \nonumber
  -u^{-2} \psi^{\prime\prime}\left(\frac{2+u+v}{2u}\right) \Bigr]\;.
\end{eqnarray}
This gives together with $U_1$ and $U_2$ all
leading-log contributions to the mobility up to order
$\epsilon^2$.

\subsection{Contribution $U_4$}\label{sec:4}
To understand the general structure of all higher-order terms,
it is very illuminating to study the next contribution. Although this is
quite tedious, it will turn out to
guide us towards an exact summation of {\em all}\, diagrams
relevant for the asymptotic low-energy behavior.
The five diagrams contributing to $U_4$ are shown in Fig.~\ref{fig5}.
Each of them can be decomposed into $4!=24$ graphs,
yielding  120 graphs in total.
Diagrams (I) to (IV) are of type (A), while diagram (V) is of type (B).
Since the calculation of $U_4$ is somewhat lengthy but proceeds
along the same lines as in Secs.~\ref{sec:22} and \ref{sec:33},
the relevant technical details are given in the
Appendix. In this section, we only discuss some conclusions
obtained from this computation.

The main findings are exhibited in Eqs.~(\ref{k14}) and (\ref{k5})
of the Appendix
which show that there is a crucial difference between diagrams
of type (A) and type (B). In type-(A) diagrams, one has a product
of only two $\sigma$'s in the integrand, while diagrams of type
(B) contain at least four $\sigma$ factors.
This characteristic behavior directs us towards an important conclusion.
Upon pulling out a factor $L_4(0)|v|/u^4$ in Eq.~(\ref{k5}), we may expand
the remaining expression in powers of $u^2$ and $|v|^2$.
Now, because of the
$\sigma$-factors, terms up to order $u^2$ and $|v|^2$ cancel out in
the sum over $\{\sigma\}$, and nonvanishing contributions from the odd time
integrations are at least
of order $u^4, |v|^4, u^2 |v|^2$. Since the asymptotic
low-energy behavior in Eq.~(\ref{leadlog}) is governed by $a_j\sim u^2,
|v|^2$, it is obvious that the asymptotic properties are
 entirely unaffected by diagram (V). It is immediately clear that also all
higher-order diagrams of type (B) do not contribute to coefficients
$a_j \sim u^2, |v|^2$. From this we conclude that, while the zero-energy
value of the conductance $G=0$ is fully determined  by the diagram
shown in Fig.~\ref{fig2},  {\em the asymptotic corrections in $u$ and $|v|$
are completely governed by diagrams of type (A)}. Following this
observation, one can make further progress. This will be discussed
in Section \ref{sec:a}.

Let us finally collect together all contributions
in order $\epsilon^3$ due to
${\rm Im}\,U_4$ for arbitrary powers in $u^2$ and $|v|^2$.
Using again an identity  of the form (\ref{identity}), the integrals
(\ref{k14}) and (\ref{k5}) given in the Appendix can be evaluated exactly.
In the end, we find
\begin{eqnarray}\label{u34}
{\rm Im}\,U_4 &=&
64\epsilon^3\,\frac{\bar\gamma}{\pi} \, L_4(0) \,{\rm Im}\Biggl[
  \psi\left(\frac{2+u+v}{2u}\right)\\
\nonumber  &-&u^{-1} \psi'\left(\frac{2+u+v}{2u}\right)
  +2 u^{-2} \psi^{\prime\prime}\left(\frac{2+u+v}{2u}\right) \\ \nonumber
  &+& u^{-3} \psi^{(3)}\left(\frac{2+u+v}{2u}\right) \Biggr]\;.
\end{eqnarray}

In view of the increasing complexity it appears to be prohibitive to go
beyond order $\epsilon^3$ in all powers of $u^2, |v|^2$. However, as the
expansion is already well advanced, in the next section
the findings obtained so far will be taken up
to conclude on exact results.
\section{RESULTS}
\subsection{Asymptotic low-energy corrections}\label{sec:a}
Our analysis in the previous section suggests that one can sum the
whole leading-log series (\ref{leadlog}) for the contributions $a_j\sim
u^2, |v|^2$, where scaled temperature
$u$ and bias $v$ are given in Eq.~(\ref{scales}). These
contributions determine the asymptotic low-energy
properties of the transport process. The leading-log
summation can indeed be performed in all orders
by considering only the asymptotic regime. As shown in the previous
section, the only diagrams contributing to this
regime are of type (A). The sum of all of them can be expressed as
$(--Y++)$, where $Y$ stands for the grand canonical gas
of finite-length dipoles of the two types $(+-)$ and $(-+)$.
(So $Y$ is basically a sum over all arrangements $X$ mentioned
in Sec.~\ref{sec:3:1}.)
In the remainder of this subsection, we will ignore all other
diagrams.\cite{foot}

The contribution $U_n^{(A)}$ to $U_n$
due to diagrams of type (A) can now be
evaluated {\em in every order} $n$.
The even time integrals result in a function $L_n(\epsilon)$ containing
the logarithm,
\begin{equation}\label{logn}
L_n(\epsilon) = \frac{1}{(n-2)!} \int_1^{\infty}
dx\, \frac{\ln^{n-2}x}{x^{1-8\epsilon}} \,e^{-|u+v|x} \;.
\end{equation}
Here, we have already integrated out all even time intervals except the
totally symmetric combination $x=x_2+\ldots+x_{2n-2}$, and we have
included the interaction factor (\ref{nnl}).
Taking into account the remaining odd time integrals and summing over
all possible arrangements of the intermediate dipoles, we find
\begin{eqnarray}
&& {\rm Im}\,\sum_{j\in (A),\,r}  U^{(j,r)}_n =  (-1)^{n-1} 
\frac{8^{n-1}}{4 u^{n-1}} L_n(\epsilon) \\ &\times& \nonumber
\int_0^\infty dx_1 \cdots dx_{2n-1}\, e^{-(
x_1+\cdots+x_{2n-1})}\\&\times&
\sum_{\{\sigma_j=\pm1\}}   \nonumber
\sigma_1 \sigma_{2n-1} \,\frac{\sin[|v|(\sigma_1 x_1
+\cdots+\sigma_{2n-1}x_{2n-1})/2]}{\sinh[u(\sigma_1 x_1
+ \cdots +  \sigma_{2n-1} x_{2n-1})/2]} \;.
\end{eqnarray}
Upon expanding the integrand, we find to lowest
order in $u$ and $|v|$ the $n$th contribution to the conductance $(n>1)$
\[
\frac{G^{(n)}}{G_0} = -2 \epsilon
 \frac{|u+v|^2}{12} \frac{(-16\epsilon)^{n-2}}{(n-2)!}
 \int_1^\infty
dx\, \frac{\ln^{n-2}x}{x^{1-8\epsilon}} \,e^{-|u+v|x} \;.
\]
Finally, summing over all $n$ gives to lowest integer orders
in $u$ and $|v|$ the expression
\begin{equation}\label{muex}
\frac{G}{G_0} =   \frac{1}{12} |u + v|^2 \left( 1 - 8\epsilon
\int_1^\infty dx \frac{e^{-|u +v|x} }{x^{1+8\epsilon}} \right) \; .
\end{equation}
An important feature should be noted. If one stops  the
expansion at the term $n=2$, an expression
of the form (\ref{muex}) would emerge
 with the factor $x^{8\epsilon}$ in the denominator of the
integrand replaced by $x^{-8\epsilon}$. The whole
nonperturbative effect of inserting a grand canonical gas of
finite-length dipoles between the charge pairs $(--)$ and $(++)$ simply
results in exchanging the factor $x^{-8\epsilon}$ in the denominator
of the integrand for the factor $x^{8\epsilon}$, as it is
already done in the integrand of (\ref{muex}). Clearly, this substitution
leads to a crucial change in the dependence on $u$ and $|v|$.

Upon performing the integral, we finally get the {\em exact asymptotic
behavior}
\begin{equation} \label{exact2}
G(V,T)/G_0 =   (1/3) \left [(eV/2 k_{\rm B} T_{\rm K})^2
+ (\pi T/T_{\rm K})^2 \right]^{1+4\epsilon}\;.
\end{equation}
This result is valid under conditions (\ref{conditions}).
For the linear conductance, the $T^{2/g-2}$
behavior found by KF is indeed recovered.
More generally, the low-temperature corrections to the zero-temperature
nonlinear conductance will follow this law as long
as $eV\ll k_{\rm B}T$. However, if $eV\gg k_{\rm B} T$,
one finds a $g$-independent universal $T^2$ law. Clearly,
for any finite value of the external voltage, the
ultimate low-temperature behavior should then be $T^2$.
The result (\ref{exact2}) describes a smooth turnover between
these two power laws and demonstrates unambiguously that
the limits $V\to 0$ and $T\to 0$ do not commute.

The nonanalytical form of the conductance is due to
the critical nature of the $(T=V=0)$ model. Inclusion
of a finite voltage breaks scale invariance, and if
one considers the temperature dependence under $V\neq 0$,
analytical behavior will be observed. The physical reason for
$T^2$ corrections is the low-frequency thermal noise
of the plasmon modes in the leads.\cite{martinis}
Similar low-temperature $T^2$ features can be found in a number of
different systems where Ohmic damping is present.\cite{weiss}
We believe that these universal $T^2$ features can indeed
be observed in the $g=\nu=1/3$ FQH transport experiments of
Milliken {\em et al.}\cite{webb} Experiments have so far been
carried out at small source-drain voltage $eV / k_{\rm B} T\approx 0.2$,
and hence the KF $T^4$ law has been found. It would be interesting
to explore the regime $eV \gg k_{\rm B} T$, where
one should be able to see the $T^2$ law. Sample heating is not
expected to cause major problems since $G$ is still quite small.
Finally we mention a completely different experimental
possibility for detection of this crossover from $T^{2/g-2}$ to
$T^2$ behaviors, namely the
novel quantum wire experiments by Tarucha {\em et al.}\cite{tarucha}

\subsection{Conjectured exact result}\label{sec:b}
Our explicit calculation for $U_n$ up to order
$n=4$ in Sec.~IV as well as the exact summation
of the leading-log series (\ref{leadlog}) to lowest
order $u^2, |v|^2$ allow us to conjecture on the asymptotically
exact result. All these results are consistent
with the following expression for the  mobility
\begin{equation}\label{exacsol}
\frac{G(V,T)}{G_0} = 1-\frac{2\gamma_{\rm r}}{eV}
\,{\rm Im}\, \psi\left(\frac{1}{2} +
 \frac{\gamma_{\rm r}+ieV/2}{2\pi k_{\rm B} T} \right)\;,
\end{equation}
which is just the exact solution (\ref{exactsol})
for $g=1/2$ with a {\em
voltage- and temperature-dependent renormalization}
\begin{equation}
\gamma \to \gamma_{\rm r} = \bar\gamma \left(
(eV/2\bar\gamma)^2
+ (\pi k_{\rm B} T/\bar\gamma)^2 \right)^{-2\epsilon}\;.
\end{equation}
We remind the reader that $\bar \gamma/k_{\rm B}$ is just the Kondo
temperature.
Expanding Eq.~(\ref{exacsol}) in powers
of $\epsilon$, one finds exactly all contributions up to order
 $\epsilon^3$ --- but in all orders of $u^2, |v|^2$ ---
as derived in Sec.~IV.
Similarly, when expanding Eq.~(\ref{exacsol}) up to order $u^2, |v|^2$, we
recover Eq.~(\ref{exact2}) in all orders of $\epsilon$.
Furthermore, the $T=0$ limit of (\ref{exacsol}) coincides with very recent
calculations by Fendley {\em et al.}\cite{fendlnew}

It is tempting, in the light of these findings, to assert that the
result (\ref{exacsol}) is the {\em exact}\, solution under conditions
(\ref{conditions}). This statement also receives support from the
correspondence of our model with an interacting fermion model.\cite{fend}
To make the
correspondence, we rewrite the solution (\ref{exacsol}) in terms of a
suitable momentum integral representation. Upon using the
representation ($x$ and $y$ real)
\begin{eqnarray*}
{\rm Im}\, \psi(1/2 +x +iy) &=&
\int\limits_0^\infty \frac{ds}{1 + s^2} \Bigl( \frac{1}{e^{2\pi(xs-y)}+1}
\\ &-& \frac{1}{e^{2\pi(xs+y)}+1} \Bigr) \;
\end{eqnarray*}
and putting $s = (v_{\rm F}^{}p/\bar\gamma)^{1+4\epsilon}$,
Eq.~(\ref{exacsol}) takes the form
\begin{eqnarray}\label{momrep}
\frac{G}{G_0} & =& 1-\frac{2v_{\rm F}^{}}{eV} \int_0^\infty
dp \,n(p)\,|S_{+-}(p)|^2 \\ &\times& \nonumber
\Bigl( f[E(p) - eV/2] - f[E(p)+eV/2]\Bigr)\; ,
\end{eqnarray}
where $f(E)$ is the Fermi function, and where
\begin{eqnarray}
|S_{+-}(p)|^2 &=& \left(1 + \left(
\frac{v_{\rm F}^{}p}{\bar\gamma}\right)^{2+8\epsilon}\right)^{-1}
\; , \label{sm}\\
n(p) &=& \left(\frac { (v_{\rm F}^{}p)^2}{ (eV/2)^2
+ (\pi k_{\rm B}^{} T)^2}\right)^{2\epsilon} \; .\label{np}\\
E(p) &=& v_{\rm F}^{} p \,n(p) \label{ep}
\end{eqnarray}
In the work by Fendley {\em et al.},\cite{fend}
a similar formula is derived for
$g=1/3$ using the thermodynamic Bethe ansatz and a Boltzmann
rate expression. The function $|S_{+-}(p)|^2$ in (\ref{sm}) agrees
with the form of the squared modulus of the exact $S$ matrix of the
quasiparticles for impurity scattering\cite{ghoshal}
on putting $g=1/2 - \epsilon$.
The function $n(p)$ represents the density of states of these quasiparticles
at voltage $V$ and temperature $T$ in the regime
(\ref{conditions}). Finally, the quantity $E(p)$ is a pseudoenergy used to
parametrize the occupation probability of the quasiparticles as $f[E(p)]$.

In conclusion, the results of our dynamical method are in full agreement with
what one would expect from combining the thermodynamic Bethe ansatz with a
Boltzmann rate expression. Moreover, we obtained the density of states and the
pseudoenergy of the quasiparticles in explicit form for $g=1/2-\epsilon$.
\section{CONCLUDING REMARKS}
We have studied transport properties of a Luttinger liquid
through a barrier at low temperature and for small voltage.
This nonperturbative problem  has been approached
by a leading-log summation technique which is valid
near the value $g=1/2$ of the interaction constant.
Putting $g=1/2-\epsilon$, the leading logs have been
summed up in all orders $\epsilon^n$ for the
asymptotic low-energy regime. Furthermore, the conductance
has been evaluated up to order $\epsilon^3$
for all powers of $V^2, T^2$.  Both findings allow to conclude on
the exact result for the low-temperature conductance near $g=1/2$.
The asymptotic low-temperature corrections exhibit a smooth
turnover from the $T^{2/g-2}$ law to a universal $T^2$
behavior as the voltage is increased. In fact, for any
finite voltage, one should find the $T^2$ law for
$k_{\rm B} T \ll eV$. We have also provided a simple physical
argument to explain the $T^2$ enhancement at finite voltage.
These $T^2$ corrections can be found even in the presence of
electron-electron backscattering.
All these Luttinger liquid fingerprints should be observable with the
fractional quantum Hall devices of Milliken {\em et al.}\cite{webb}
We predict that application of
a source-drain voltage drop $eV\gg k_{\rm B} T$ will
change the KF $T^4$ law into a $T^2$ behavior.

To conclude, we mention that our method provides the unique possibility
of performing exact noise calculations. It seems very difficult or even
impossible to obtain ac noise properties with any other technique
available at the moment.

\acknowledgements
This work was partially supported by the EC SCIENCE program.
We have benefitted from discussions with P. Fendley,
L.I. Glazman, H. Grabert, F. Guinea, C.H. Mak, H. Saleur
and A.D. Zaikin.

\appendix
\section*{Contributions to $U_4$}
\label{sec:appendix}

This appendix contains the technical details of the computation of
the contribution $U_4$ to the conductance discussed in Sec.~\ref{sec:4}.
The $96$ graphs related to the diagrams (I) to (IV)
shown in Fig.~\ref{fig5} fall into four classes
regarding their dependence on the long even  times $x_2,x_4,x_6$.
Denoting combinations of the short odd times as $a,b,c,d$, these
four classes are given by
\begin{eqnarray*}
f^{(1)}_4& =&
 \Bigl( \sinh[u(x_2+a)/2]\sinh[u(x_4+b)/2]\\ &\times& \sinh[u(x_6+c)/2]
\sinh[u(x_2+x_4+x_6+d)/2]\Bigr)^{-1}\\
f^{(2)}_4& =&
 \Bigl( \sinh[u(x_2+a)/2]\sinh[u(x_6+b)/2]\\ &\times& \sinh[u(x_2+x_4+c)/2]
\sinh[u(x_4+x_6+d)/2]\Bigr)^{-1}\\
f^{(3)}_4& =&
 \Bigl( \sinh[u(x_4+a)/2]\sinh[u(x_4+b)/2]\\ &\times&
\sinh[u(x_2+x_4+x_6+c)/2] \\ &\times&
\sinh[u(x_2+x_4+x_6+d)/2]\Bigr)^{-1}\\
f^{(4)}_4& =&
 \Bigl( \sinh[u(x_4+a)/2]\sinh[u(x_2+x_4+b)/2]
\\ &\times& \sinh[u(x_4+x_6+c)/2] \sinh[u(x_2+x_4+x_6+d)/2]\Bigr)^{-1}\; .
\end{eqnarray*}
Putting $N^{(A)}_4 =1$ for the moment
in analogy to Sec.~\ref{sec:33}, the triple integral
\begin{equation}\label{f4}
\int_1^\infty dx_2 dx_4 dx_6 \,e^{-v(x_2+x_4+x_6)}
\, f^{(r)}_4
\end{equation}
must give the logarithms, and we are interested in terms $\sim
\ln^3|u +v|$.
It is straightforward to show that the class $f_4^{(3)}$ does
not give leading logs. The other three classes, however, give
leading logarithmic contributions.

Switching to relative times and to the variable $x = x_2+x_4+x_6$, the
former are easily integrated out. Upon performing the shift $x + q \to x$,
where $q$ is $a$, $b$, $c$, or $d$, and restoring
$N^{(A)}_4 = x^{8\epsilon}$, the remaining integral resulting
from Eq.~(\ref{f4}) is
\begin{equation}\label{g40}
L_4(\epsilon) = \frac{1}{2!} \int_1^{\infty} dx\, \frac{\ln^2 x}{
x^{1-8\epsilon}} \,e^{-|u+v|x} \; .
\end{equation}
Having evaluated the even time integrals in this way, the remaining odd
time integrals of the relevant graphs take forms such that the sum of
them indeed gives an expression fitting to the previous forms
(\ref{a2}) and (\ref{kkk}),
\begin{eqnarray} \nonumber
&&{\rm Im}\,\sum_{j\in (A),\,r} U^{(j,r)}_4 = - \frac{128}{u^3} L_4(\epsilon)
\int_0^\infty dx_1 dx_3 dx_5 dx_7\\ &\times& e^{-
(x_1+x_3+x_5+x_7)}\label{k14}
 \sum_{\{\sigma=\pm1\}}
\sigma_1 \sigma_7 \\ &\times& \frac{\sin[|v|(\sigma_1 x_1
+ \sigma_3 x_3 + \sigma_5 x_5+\sigma_7 x_7)/2]}{\sinh[u(\sigma_1
x_1
+ \sigma_3 x_3 + \sigma_5 x_5+ \sigma_7 x_7)/2]} \;. \nonumber
\end{eqnarray}
This contains all contributions of graphs due to the type-(A)
diagrams (I) to (IV).
Using the same procedure as before for $U_3$, this expression can be
evaluated without any further
approximation. Before stating  the result, we
discuss the remaining diagram (V) in Fig.~\ref{fig5}.

The 24 graphs related to diagram (V) fall into three classes.
One is just $f_4^{(3)}$ from above and does not contribute to leading
logarithms. Another one is $f_4^{(4)}$ and does contribute. The third
and last type is of the form
\begin{eqnarray*}
&& \Bigl( \sinh[u(x_2+x_4+a)/2]\sinh[u(x_2+x_4+b)/2]
\\ &\times& \sinh[u(x_4+x_6+c)/2]
\sinh[u(x_4+x_6+d)/2]\Bigr)^{-1}
\end{eqnarray*}
and does not contribute to leading logs.
In view of the before-mentioned complications of the even time integrations
due to the residual interaction factor $N^{(B)}_4$,
we restrict our attention to the order $\epsilon^3$ in the diagram (V),
i.e.~we put $N^{(B)}_4 = 1$ and
do not push the explicit calculation  beyond this order.
Carrying out the even time integrations as before,
we then find the factor $L_4(0)$.

The remaining odd time integrals of the relevant graphs can be written
in such a form that the contribution of diagram (V) emerges as
\begin{eqnarray} \nonumber
&&{\rm Im} \,\sum_{j\in (B),\,r} U^{(j,r)}_4 = \frac{64}{u^3} L_4(0)
\int_0^\infty dx_1 dx_3 dx_5 dx_7\\ &\times& e^{-
(x_1+x_3+x_5+x_7)}\label{k5}
 \sum_{\{\sigma_j=\pm1\}}
\sigma_1 \sigma_3 \sigma_5 \sigma_7\\ &\times&\frac{\sin[|v|(\sigma_1 x_1
+ \sigma_3 x_3 + \sigma_5 x_5+\sigma_7 x_7)/2]}{\sinh[u(\sigma_1
x_1 + \sigma_3 x_3 + \sigma_5 x_5+ \sigma_7 x_7)/2]} \;.
\nonumber
\end{eqnarray}
The crucial difference to the expression (\ref{k14}) for diagrams  (I)
to (IV) is the sign factor $\sigma_1 \sigma_3 \sigma_5 \sigma_7$
instead of $\sigma_1 \sigma_7$. These distinguishing features are easily
extended to higher-order diagrams
 $n>4$ as follows. Every {\em odd}\, time interval
$\tau_{2j-1}$ that is fenced by charges of the same sign gives a factor
$\sigma^{}_{2j-1}$ in the integrands of the respective functions
${\rm Im}\,\sum_{j\in (A),\,r} U^{(j,r)}_n$ and
${\rm Im}\,\sum_{j\in (B),\,r} U^{(j,r)}_n$, see Eqs.~(\ref{k14}) and
(\ref{k5}). Diagrams of type (A) involve a product of only
two $\sigma^\prime s$ in the integrand, while diagrams of type (B)
imply a product of at least four $\sigma^\prime s$.
The final result for $U_4$ and the consequences
of this observation are discussed in detail in Sec.~\ref{sec:4}.


\begin{references}
\bibitem{footn} Note that their contributions cancel each
other when they are collapsed as these dipoles are inserted into even time
intervals.
\bibitem{foot} A systematic computation of
the next-leading contributions $a_j\sim u^4,|v|^4,u^2|v|^2$ could be
feasible
by noticing that these are completely determined by diagrams of
type $(--Y--Y++Y++)$ in addition to $(--Y++)$.
\end{references}
\end{document}